\documentclass[reprint,superscriptaddress,showpacs,amsmath,amssymb,prb]{revtex4-2}
\usepackage{natbib}
\usepackage{siunitx}
\usepackage{array}
\usepackage{xcolor}
\usepackage{subfigure} 
\usepackage{amsmath} 
\usepackage{amsfonts}
\usepackage{amssymb}
\usepackage{graphicx}
\usepackage{hyperref}
\usepackage{appendix}
\usepackage{academicons}
\usepackage{orcidlink}

\newcommand{\orcid}[1]{\href{https://orcid.org/#1}{\textcolor[HTML]{A6CE39}{\aiOrcid}}}
\providecommand{\vect}[1]{{\boldsymbol{#1}}}

\usepackage{physics}

\begin{document} 
\title{Amplifying the antiferromagnetic spin Seebeck effect through topological magnons}

\author{Feodor Svetlanov Konomaev} 
\affiliation{Department of Engineering Sciences, University of Agder, 4879 Grimstad, Norway} 
\author{Kjetil M. D. Hals}
\affiliation{Department of Engineering Sciences, University of Agder, 4879 Grimstad, Norway} 
\date{\today}
\newcommand{\Kjetil}[1]{\textcolor{red}{#1}} 
\begin{abstract}
Topological magnons emerge as topologically protected spin wave states at the edges of magnets. Here, we theoretically explore how these surface states can be harnessed to amplify the spin Seebeck effect (SSE) in antiferromagnets (AFMs) interfaced with normal metals (NMs). Based on a microscopic model of a kagome AFM, we demonstrate that broken mirror symmetry, combined with the Dzyaloshinskii-Moriya interaction (DMI), drives the system into a topological phase hosting spin-polarized magnons at the boundaries. Notably, linear response calculations reveal that in AFM/NM heterostructures, the topological magnons exhibit strong coupling to the metal’s charge carriers, resulting in a substantial enhancement of the SSE. The relative contribution of the topological magnons is found to be 4-5 times greater than that of the trivial magnon bands. Moreover, our results show that this enhancement is highly sensitive to the strength of the DMI.
\end{abstract}

\maketitle 

\section{Introduction} 
The spin-Seebeck effect (SSE) refers to the generation of spin currents in magnetic systems due to temperature gradients~\cite{Johnson:PRB1987, Hatami:PRL2007, Uchida:Nat2008, Uchida:NatMat2010, Jaworski:NatMat2010, Uchida:APL2010, Guo:PRX2016}. In magnetic insulators, these spin currents are carried by magnons -- the elementary spin excitations of the ordered spin structure. 
A key implication of the SSE is its ability to convert heat into electricity through the spin degree of freedom. This heat-to-electricity conversion was first demonstrated in heterostructures composed of ferromagnetic insulators (FIs) and heavy metals (HMs)~\cite{Uchida:Nat2008, Uchida:NatMat2010, Jaworski:NatMat2010, Uchida:APL2010, Adachi:prb2011, Kirihara:NatMat2012, Guo:PRX2016}. Here, a temperature gradient across the FI/HM interface generates a magnon-driven spin current that flows into the HM layer. This spin current is further converted into electricity via the inverse spin Hall effect. Later, a similar thermoelectric effect was observed in heterostructures based on antiferromagnets (AFMs)~\cite{Ohnuma:prb2013, Seki:PRL2015, Wu:PRL2016, Shiomi:PRB2017, Shi:PRL2019, Ross:PRB2021, Xu:PRL2022}, with an SSE coefficient reported to be fifty times larger than that of FI/HM structures~\cite{Wu:PRL2016}. When the magnetic heterostructures are applied to hot surfaces, they function as unique thermoelectric generators~\cite{Kirihara:NatMat2012} that offer clear advantages over traditional thermoelectric technologies. First, they exhibit excellent scalability, where the power output can be increased by expanding the area of the thin-film heterostructure. Second, heterostructures are considerably simpler to fabricate than conventional thermocouple-based modules. However, despite these promising features, the efficiency of the spin-based generators remains too low for large-scale commercial applications~\cite{Uchidal:Review}.

\begin{figure}[ht] 
\centering 
\includegraphics[scale=1.0]{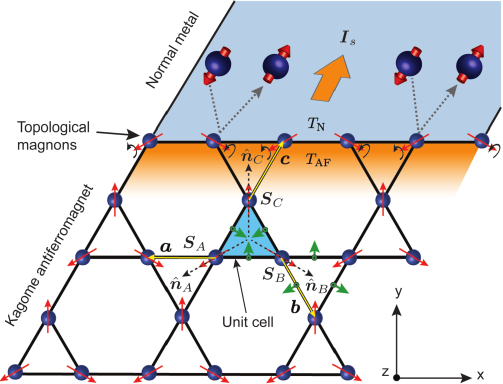}  
\caption{(color online). A heterostructure composed of a kagome AFM interfaced with an NM. Topological magnons, localized at the edge of the AFM, strongly couple to the charge carriers of the NM, resulting in an enhancement of the thermally-driven spin current pumped into the NM. The green arrows represent the DMI vectors along the various bonds.}
\label{Fig1} 
\end{figure} 

An intriguing and largely unexplored mechanism for enhancing the thermoelectric efficiency of magnetic heterostructures is the presence of topological magnon states. These states emerge in both antiferromagnetic and ferromagnetic systems with magnon bands characterized by nonzero Chern numbers~\cite{Zhang:PRB2013, Shindou:PRB2013A, Shindou:PRB2013B, Mook:PRB2014A, Mook:PRB2014B, Mook:PRB2015, Romhanyi:PRB2016, Mook:PRB2016, Mook:PRL2016, Molina:NJP2016, Li:NatCom2016, Owerre:2016, Laurell:PRB2018, McClarty:PRB2018, Mook:PRB2019, Dias:PRL2019, Bhowmick:PRB2020, Corticelli:PRL2023, McClarty:Review}. The topological properties can be induced through periodically modulated magnetization and exchange interactions or via the Dzyaloshinskii–Moriya interaction (DMI) produced by the intrinsic spin-orbit coupling (SOC).
In the topologically nontrivial phase, the topological magnons manifest as coherent, propagating spin waves along the sample's surface, offering energy-efficient channels for transporting spin currents. The topological spin excitations can be experimentally observed through the magnon thermal Hall effect~\cite{Laurell:PRB2018, Mook:PRB2019, Onose:Science2023, Ideue:PRB2012, Hirschberger:PRL2015}. 
Additionally, recent inelastic neutron scattering experiments have uncovered signatures of topologically nontrivial magnon states in both three-dimensional (3D) and two-dimensional (2D) ferromagnetic~\cite{Chisnell:PRL2015, Chen:PRX2018, Yuan:PRX2020, Elliot:NatComm2021, Scheie:PRL2022} and antiferromagnetic systems~\cite{Yao:NatPhys2018,Bao:NatComm2018}.

Since the topological magnons are localized at the boundaries, they are expected to strongly couple to the itinerant charge carriers of a metal interfaced with the magnetic insulator. This strong coupling can notably amplify interface phenomena, such as the SSE, significantly improving the overall heat-to-electricity conversion efficiency of the magnetic heterostructures. 

In this work, we theoretically explore how spin-polarized topological manginess enhance the SSE in a kagome AFM interfaced with a normal metal (NM). 
We demonstrate that the combination of broken mirror symmetry and a finite in-plane DMI is crucial for inducing both a net spin polarization of the magnons and driving the AFM into a topological phase.
Additionally, we present a comprehensive microscopic theory for calculating the SSE in heterostructures composed of coupled layers of kagome AFMs and NMs. Using this framework, we show that the topologically protected edge states strongly couple to the charge carriers at the AFM/NM interface, leading to a substantial enhancement of the SSE. Specifically, the contribution of the topological magnons to the spin current is typically 4–5 times greater than that of trivial states.
To date, few experiments have investigated the SSE in noncollinear AFMs~\cite{Xu:PRL2022}, such as kagome AFMs. Our findings show that these largely unexplored spin systems exhibit a distinct topological SSE, which offers a promising mechanism for optimizing the performance of spin-based thermoelectric generators.

\section{Theory} 
The kagome AFM is described by the Hamiltonian~\cite{Rodrigues:PRB2022}
\begin{equation}
\mathcal{H}_\text{AF}=\mathcal{H}_\text{E}+\mathcal{H}_\mathrm{A}+\mathcal{H}_\mathrm{DM}\text{.}
 \label{eq:ham}
\end{equation}
Here, $\mathcal{H}_\text{E}=J\sum_{\langle ij\rangle} \vect{S}_i\cdot\vect{S}_j$ represents the exchange interaction between the nearest neighbour spins $\vect{S}_i$ and $\vect{S}_j$ with the coupling strength $J>0$, $\mathcal{H}_\mathrm{A}=\sum_i \left(K_\perp(\vect{S}_i\cdot\hat{\vect{z}})^2-K(\vect{S}_i\cdot\hat{\vect{n}}_i)^2\right)$ characterizes the easy plane ($K_\perp>0$) and easy axes ($K>0$) anisotropy energies, respectively, and $\mathcal{H}_\mathrm{DM}=\sum_{\langle ij\rangle}\vect{D}_{ij}\cdot(\vect{S}_i\times\vect{S}_j)$ is the DMI between the nearest neighbor lattice sites $\langle ij\rangle$. The vectors $\hat{\vect{n}}_i$ define the local easy axis at each lattice point $i$. In  kagome AFMs, the three sublattice spins within a unit cell experience distinct easy axes, denoted as 
$\hat{\vect{n}}_A=[-\frac{\sqrt{3}}{2},-\frac{1}{2},0]^T$,  $\hat{\vect{n}}_B=[\frac{\sqrt{3}}{2},-\frac{1}{2},0]^T$ and $\hat{\vect{n}}_C=[0,1,0]^T$ (see Fig.~\ref{Fig1}). 
The DMI vectors $\vect{D}_{ij}$ are determined by the symmetry of the system. In a kagome AFM with broken mirror symmetry of the lattice plane, the DMI vectors connecting the sites within a single unit cell are given by~\cite{Rodrigues:PRB2022}
\begin{equation}\label{eq:d-s}
    \begin{split}
        & \vect{D}_{AB}=D_\perp\hat{\vect{z}}+D_\parallel (\hat{\vect{a}}\times\hat{\vect{z}} )\text{,}  \\
& \vect{D}_{BC}=D_\perp\hat{\vect{z}}+D_\parallel (\hat{\vect{b}}\times\hat{\vect{z}} )\text{,} \\
& \vect{D}_{CA}=D_\perp\hat{\vect{z}}+D_\parallel (\hat{\vect{c}}\times\hat{\vect{z}} )\text{,}
\end{split}
    \end{equation}
where $\vect{a}=a[ -1,0,0]^T$ is the vector that connects site $B$ to site $A$, $\vect{b}=a[\frac{1}{2},-\frac{\sqrt{3}}{2},0]^T$ connects site $C$ to site $B$, and $\vect{c}=a[\frac{1}{2},\frac{\sqrt{3}}{2},0]^T$ 
is connecting site $A$ to site $C$ with $a$ being the lattice constant (see Fig.~\ref{Fig1}). The DMI vectors on the bondings connecting different unit cells are determined by inversion about site $B$~\cite{Rodrigues:PRB2022}.

Due to the parallel component of the DMI, the spins acquire an out-of-plane tilting by an angle 
\begin{equation}
\begin{split}
\theta=-\frac{1}{2}\arctan\left( \frac{2\sqrt{3}D_\parallel}{3J+K_\perp+K-\sqrt{3}D_\perp}\right) \text{,}
\end{split}
 \label{eq:thetaI}
\end{equation}
which is found by minimizing the energy of Eq.~\eqref{eq:ham} (see the Appendix~\ref{ap:theta}). 
Consequently, in equilibrium, the sublattice spins are given by $\vect{S}_i^{(0)} = \hbar S [ \cos(\theta)\hat{\vect{n}}_i+\sin(\theta)\hat{\vect{z}} ]$ for $i\in\{A,B,C\}$.
For a mirror symmetric kagome lattice plane, the in-plane component $D_\parallel$ vanishes, leaving only the out-of-plane component $D_\perp$. 
In this case, the spins lie in $xy$-plane with $\theta=0$.

Hereafter, we consider a system defined by the material parameters $J=10$ meV/$\hbar^2$, $K=0.03$ meV/$\hbar^2$, $K_\perp=0.9$ meV/$\hbar^2$, $D_\perp=-0.2$ meV/$\hbar^2$, $S=1$, and $a=3$~\AA,
which are representative of kagome AFMs~\cite{Rodrigues:PRL2021}. We allow $D_\parallel$ to vary in order to highlight the crucial role this parameter plays in enabling topological states 
and generating spin-polarized magnons, as discussed below. Notable candidate materials described by the above spin model include Mn$_3$X (X= Ga, Ge, Sn), iron jarosites, and rare-earth kagome compounds.
As we will demonstrate below, achieving topological magnon bands with a global gap requires a relatively large in-plane DMI of $D_\parallel /J > 0.304$. 
Such a large DMI has been experimentally observed in the rare-earth kagome compound Nd$_3$Sb$_3$Mg$_2$O$_{14}$, where they estimate the $D_\parallel /J $ ratio to be 0.8~\cite{Scheie:PRB2016}. 
Moreover, it is possible to significantly enhance the in-plane DMI in more stable materials, such as Mn$_3$X, by growing a monolayer on a substrate that breaks the mirror symmetry of the kagome lattice plane.

To describe the collective spin excitations of the Hamiltonian~\eqref{eq:ham}, we employ the Holstein-Primakoff transformation~\cite{Auerbach:book}, which expresses the spin operators in terms of the bosonic ladder operators $a_i$ and $a_i^\dagger$:
$S_{i,x_i}=\frac{\hbar\sqrt{2S}}{2}(a_i+a_i^\dagger)$, $S_{i,y_i}=\frac{\hbar\sqrt{2S}}{2i}(a_i-a_i^\dagger)$, and $S_{i,z_i}=\hbar(S-a_i^\dagger a_i)$.
Here, $\hat{\vect{x}}_i= \hat{\vect{n}}_i\cross\hat{\vect{z}}$, $\hat{\vect{y}}_i= \hat{\vect{z}}_i\cross\hat{\vect{x}}_i$, and $\hat{\vect{z}}_i=\vect{S}_i^{(0)}/\hbar S$ define the local reference frame at lattice site $i$ such that the quantization axis $\hat{\vect{z}}_i$ points along $\vect{S}_i^{(0)}$. The expressions for the spin operators are substituted into Eq.~\eqref{eq:ham} followed by the Fourier transformation $a_i = (1/\sqrt{N_{\rm uc}})\sum_{\vect{k}} a_{l, \vect{k}} \exp (i(\vect{R}_j + \vect{\delta}_l )\cdot\vect{k})$, where $N_{\rm uc}$ is the number of magnetic unit cells, and $\vect{R}_j + \vect{\delta}_l$ denotes the position of spin $i$, expressed in terms of the unit cell position vector $\vect{R}_j$ and the vector $\vect{\delta}_l$ (with $l\in\{A,B,C\}$) specifying the position of sublattice spin $l$ within unit cell $j$. 
The above substitution and transformation map the spin Hamiltonian~\eqref{eq:ham} onto a  bosonic Bogoliubov-de-Gennes (BdG) Hamiltonian for the ladder operators
\begin{equation}\label{eq:bdg}
    \mathcal{H}_\text{AF}=\frac{1}{2}\sum_{\vect{k}} \vect{\Psi}_{\vect{k}}^{\dagger} \mathbf{H}_{\vect{k}} \vect{\Psi}_{\vect{k}}
    \text{,}
    \end{equation}
where $\vect{\Psi}_{\vect{k}}^{\dagger}= [\vect{\alpha}_{\vect{k}}^{\dagger}, \vect{\alpha}_{-\vect{k}} ]$ with $\vect{\alpha}_{\vect{k}}^\dagger=[a_{A,\vect{k}}^\dagger,a_{B,\vect{k}}^\dagger,a_{C,\vect{k}}^\dagger]$.
Here, and in the following expressions,  $\vect{k}$ denotes wave vectors within the first Brillouin zone. The full $6\times 6$ matrix form of $\mathbf{H}_{\vect{k}}$ is provided in Appendix~\ref{app:bdg}.

The BdG Hamiltonian~\eqref{eq:bdg} is diagonalized by a  
$6\times6$ paraunitary matrix $\mathbf{T}_{\vect{k}}^{-1}= [ \mathbf{U}_{\vect{k}} , \mathbf{V}_{-\vect{k}}^* ;   \mathbf{V}_{\vect{k}} , \mathbf{U}_{-\vect{k}}^*] $ where $\mathbf{U}_{\vect{k}} $ and $\mathbf{V}_{\vect{k}} $ are $3 \times3$ submatrices~\cite{Colpa}: 
\begin{equation}\label{eq:transform}
    (\mathbf{T}_{\vect{k}}^{-1})^\dagger \mathbf{H}_{\vect{k}}\mathbf{T}_{\vect{k}}^{-1}=\mathrm{diag}(\mathbf{E}_{\vect{k}},\mathbf{E}_{-\vect{k}})\text{.}
    \end{equation}
In the above expression, $\mathbf{E}_{\vect{k}}=\mathrm{diag}\left(\varepsilon_3(\vect{k}),\varepsilon_2(\vect{k}),\varepsilon_1(\vect{k})\right)$ with $\varepsilon_n (\vect{k})$ being the eigenenergies of the $n$th band.
The diagonalized form of Eq.~\eqref{eq:bdg} is $\mathcal{H}=\sum_{n,\vect{k}} \varepsilon_n (\vect{k}) \gamma^{\dagger}_{n,\vect{k}} \gamma_{n,\vect{k} }$ where  $\gamma^{\dagger}_{n,\vect{k}}$ and $\gamma_{n,\vect{k} }$
are the ladder operators describing the elementary magnonic spin excitations of the AFM.
The Hamiltonian~\eqref{eq:bdg} is diagonalized numerically~\cite{Colpa}. 

\begin{figure}[ht] 
\centering 
\includegraphics[scale=1.0]{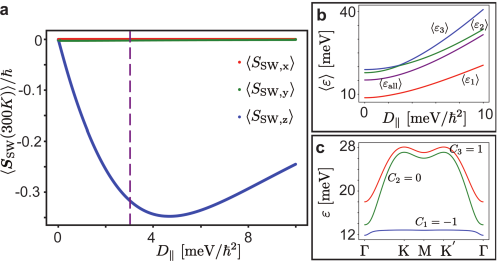}  
\caption{(color online). {\bf a}. The average spin polarization of the magnons at 300 K as function of the in-plane DMI. Dotted line indicates the transition into the topological regime with Chern numbers $C_1=-1$, $C_2=0$, and $C_3=1$. {\bf b}. The average energy of the three magnon bands as a function of the in-plane DMI. {\bf c}. The magnon energy bands along high-symmetry lines in the Brillouin zone for $D_\parallel=5$~meV/$\hbar^2$.}
\label{Fig2} 
\end{figure} 

\section{Results and discussion}
The thermally excited magnons must, on average, carry a net spin angular momentum in order to pump a DC spin current into an adjacent metallic layer. 
Thus, we begin by examining the spin polarization of the magnons. 

The statistical average of the magnons' spin polarization at temperature $T$ is given by
\begin{equation}\label{eq:spinpolT}
\langle\vect{S}_{\mathrm{SW}}(T)\rangle=\sum_{n,\vect{k}}\langle\vect{S}_{\mathrm{SW};n, \vect{k}}\rangle n_\mathrm{B}(\varepsilon_n(\vect{k}),T)\text{,}
\end{equation}
where $n_\mathrm{B}(\varepsilon,T)=(e^{\varepsilon/k_BT}-1)^{-1}$ is the Bose-Einstein distribution and
the quantity  $\langle\vect{S}_{\mathrm{SW};n, \vect{k} }\rangle = -[ \langle n,\vect{k} | \sum_i \vect{S}_i | n,\vect{k} \rangle - \langle \vect{S}_{\rm vac} \rangle  ]$ represents the average spin carried by the magnon mode $| n,\vect{k} \rangle = \gamma_{n,\vect{k}}^{\dagger} | 0 \rangle$. Here, $\langle \vect{S}_{\rm vac} \rangle$ is the spin of the the vacuum state $| 0 \rangle$.  
For further analysis, it is convenient to express $\langle\vect{S}_{\mathrm{SW};n, \vect{k}}\rangle$ in terms of the elements of the paraunitary matrix $\mathbf{T}_{\vect{k}}^{-1}$ (see Appendix~\ref{app:sp}):
\begin{equation}\label{eq:spinpol}
\langle\vect{S}_{\mathrm{SW}; n, \vect{k}}\rangle=\hbar\sum_{l}(\abs{U_{ l n, \vect{k}}}^2+\abs{V_{l n, \vect{k}}}^2)\hat{\vect{z}}_l\text{.}
\end{equation}

Fig.~\ref{Fig2}a shows the $x$, $y$ and $z$ components of the average spin polarization$\langle\vect{S}_{\mathrm{SW}}(T)\rangle$  as a function of $D_\parallel$  for a kagome AFM at a temperature of $T=300$ K.
Notably, the magnon system exhibits a net spin polarization along the z-axis when $D_\parallel \neq 0$. This polarization arises from the out-of-plane tilting in Eq.~\eqref{eq:thetaI}, which results in a weak ferromagnetic phase and a corresponding net polarization of the thermally excited magnons. Additionally, we observe that the spin polarization reaches its peak value for a finite in-plane DMI .
The reason the polarization does not increase monotonically with $D_\parallel$
is that the in-plane DMI also elevates the magnon state energies. This is depicted in the inset of Fig.~\ref{Fig2}b, which shows the average energy of the first, second, and third bands, and the average energy of all bands, for different $D_\parallel$. Thus, $D_\parallel$ induces two counteracting effects that influence the polarization: (1) the increased tilting angle, which enhances $\langle\vect{S}_{\mathrm{SW}; n, \vect{k}}\rangle $  for each magnon state, and (2) the rise in magnon energy, which reduces the Bose-Einstein occupation number $n_\mathrm{B}(\varepsilon,T)$, leading to a lower overall $\langle\vect{S}_{\mathrm{SW}}(T)\rangle$. The maximum spin polarization occurs at the $D_\parallel$ -value where these two effects are balanced.

In addition to generating spin-polarized magnons, the in-plane DMI is pivotal in driving the spin system into a topological phase characterized by magnonic edge states. We investigate this transition by numerically calculating the Chern numbers for the three magnon bands, expressed as the normalized integrals of the Berry curvatures $B_n(\vect{k})$ ($n\in \{ 1,2,3 \}$) over the first Brillouin zone:
\begin{equation}\label{eq:chern}
    C_n=\frac{1}{2\pi}\int d\vect{k} B_n(\vect{k})\text{.}
    \end{equation}
 Here, $ B_n(\vect{k})=\epsilon_{\mu\nu}\partial_{k_\mu}A_{n,\nu}(\vect{k})$, where  
 $ A_{n,\nu}(\vect{k})=i[\boldsymbol{\sigma}_3(\mathbf{T}_{\vect{k}}^{-1})^\dagger\boldsymbol{\sigma}_3\partial_{k_\nu}\mathbf{T}_{\vect{k}}^{-1}]_{nn}$ ($\nu\in\{x,y\}$) is the gauge connection,
 $\boldsymbol{\sigma}_3=\mathrm{diag}(1,1,1,-1,-1,-1)$, and $\varepsilon_{\mu\nu}$ denotes the 2D antisymmetric Levi-Civita symbol~\cite{Shindou:PRB2013B}.
 
Our numerical analysis shows that an in-plane DMI exceeding 3.04 meV is required to obtain well-defined Chern numbers with a global gap between the two lowest magnon bands, as illustrated in Fig.~\ref{Fig2}c for $D_\parallel = 5$.
In this regime, the Chern numbers are $C_1=-1$, $C_2=0$, and $C_3=1$, signifying the presence of a single edge state band within the global gap between the two lowest magnon bands for a finite system with boundaries.      
Additionally, in a finite system, an edge state band forms between the second and third bands. However, this band does not reside within a global gap (see Fig.~\ref{Fig2}c).
For $D_\parallel < 3.04$~meV, narrow regions in parameter space exhibit nonzero Chern numbers (different from those mentioned above), but without a global gap between neighboring magnon bands. 
Consequently, in what follows, we focus on the regime where $D_\parallel > 3.04$~meV, as this ensures a clear identification of the contribution of the nontrivial edge states to the SSE.

\begin{figure}[ht] 
\centering 
\includegraphics[scale=1.0]{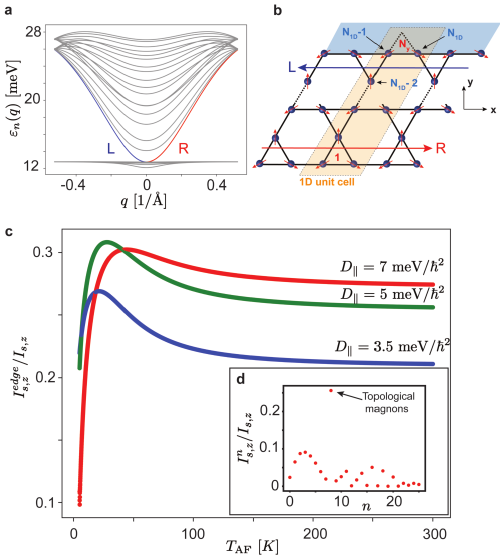}  
\caption{(color online). {\bf a}. Magnon bands of a quasi-1D system with $D_\parallel=5$~meV/$\hbar^2$. Right- (left-) propagating topological magnons are highlighted in red (blue). {\bf b}. The unit cell of the quasi-1D system. {\bf c}. Relative contribution of first-gap edge states to the total spin current. {\bf d}. Relative contribution of all bands $\varepsilon_n$ to the total spin current at 300 K for $D_\parallel=5$~meV/$\hbar^2$.
In all computations, we have used $N_y=9$.}
\label{Fig3} 
\end{figure}

Next, we explore how the surface states influence the SSE.    
To this end, we consider a quasi-1D kagome AFM that extends infinitely along the $x$-axis, with a finite thickness of $N_y$ primitive magnetic unit cells in the $y$-direction, terminating in flat edges. This quasi-1D system is defined by a new composite unit cell comprising $N_{1D}=3N_y-1$ sites (Fig.~\ref{Fig3}b). The 1D AFM strip is further connected to a 2D NM sample.
Similar magnetic heterostructure configurations have been employed to investigate quasiparticle properties in magnetic systems~\cite{Wang:arXiv24, Joshi:prb2018}.
As our primary focus is to investigate the contribution of the topological magnons relative to the trivial states, we disregard external magnetic fields~\cite{Comment2}. 
The Hamiltonian governing the AFM/NM heterostructure is given by:
\begin{equation}
\mathcal{H}=\mathcal{H}_\text{AF}^{\rm (1D)}+\mathcal{H}_\mathrm{NM}+\mathcal{H}_\mathrm{I}\text{.}
 \label{eq:hamAF_NM}
 \end{equation}
Here, $\mathcal{H}_\text{AF}^{\rm (1D)}= (1/2)\sum_{q} \vect{\Psi}_{q}^{\dagger} \mathbf{H}_{q} \vect{\Psi}_{q}$ where $ \mathbf{H}_{q}$ is
the $2N_{1D}\times 2N_{1D}$ BdG Hamiltonian characterizing the 1D AFM, which is derived following the same steps as in the derivation of Eq.~\eqref{eq:bdg} with $\vect{k} = q \hat{\vect{x}}$ and $\vect{\delta}_l$ running over the $N_{1D}$ sites within the 1D magnetic unit cell.   
$\mathcal{H}_\mathrm{NM}=\sum_{\vect{k},\vect{q}}\mathbf{c}_{\vect{k}}^\dagger \left(\varepsilon_{\vect{k}}\delta_{\vect{k},\vect{q}}+U_{\vect{k}-\vect{q}}\left(1+i \alpha_\mathrm{so}\vect{\sigma}\cdot(\vect{k}\times\vect{q})\right)\right)\mathbf{c}_{\vect{q}}$
describes the itinerant charge carriers of the NM, where $\mathbf{c}_{\vect{k}}^{\dagger} = [c_{\vect{k},\uparrow}^{\dagger} , c_{\vect{k},\downarrow}^{\dagger}]$ with
  $c_{i\tau}^{\dagger}$ ($c_{i\tau}$) being the fermionic creation (destruction) operator, $U_{\vect{k}-\vect{q}}$ is the Fourier transform of the impurity potential, $\alpha_{\rm so}$ parametrizes the SOC strength at the impurities, and $\varepsilon_{\vect{k}}$ represents the energy of the free electrons. $\mathcal{H}_\mathrm{I}= \sum_{i\in I} J_{sd} \vect{s}_i\cdot \vect{S}_i$ expresses the interfacial exchange interaction between the AFM's spins $\vect{S}_i$ localized  at the AFM/NM interface ($I$) and the charge carriers' spin-density $\vect{s}_i = (\hbar/2) c_{i\tau}^{\dagger} \boldsymbol{\sigma}_{\tau\tau^{'}} c_{i\tau^{'}}$. Here, $\boldsymbol{\sigma}$ is a vector consisting of the Pauli matrices. 
  
Numerical diagonalization of $\mathcal{H}_\text{AF}^{\rm (1D)}$ yields the magnon bands of the quasi-1D system. Fig.~\ref{Fig3}a presents the dispersion relation for a kagome AFM strip with a thickness of $N_y=9$ primitive unit cells. As expected, edge states emerge in the first gap, with the right-moving edge states highlighted in red and the left-moving ones in blue. The edge states in the second gap cannot be identified due to the intermixing of the second and third bands.

The Heisenberg equation $\dot{\vect{s}}_{\rm Tot}= (i/\hbar) [\mathcal{H}_{\rm I}, \vect{s}_{\rm Tot} ] $, with $\vect{s}_{\rm Tot}= \sum_{i} \vect{s}_i$, describes the rate of change of the total spin in the NM due to its coupling to the AFM. 
The DC spin current $\vect{I}_\mathrm{s}$ pumped into the NM can thus be obtained by taking the statistical average of this equation:
$\vect{I}_\mathrm{s}=  (i/\hbar) \langle [\mathcal{H}_{\rm I}, \vect{s}_{\rm Tot} ]  \rangle $.
The average is evaluated by treating the interfacial exchange coupling $\mathcal{H}_\mathrm{I}$ as a perturbation and applying linear response theory. 
In the limit of weak damping in the AFM, the final result becomes
\begin{equation}\label{eq:ultimate}
\vect{I}_\mathrm{s}= \mathcal{K}\sum_{n, q }\vect{\Omega}_{n;q}f_{n;q}\left( \coth\frac{\varepsilon_n(q)}{2k_BT_\mathrm{N}} - \coth\frac{\varepsilon_n(q)}{2k_BT_\mathrm{AF}} \right) ,
\end{equation}
where $n$ runs over all magnon energy states $\varepsilon_n(q)=\hbar \omega_n (q)$ for the wave vector $q$ along the $x$-axis. (see Appendix~\ref{app:model})
The parameter $\mathcal{K}=(L_y\chi_\mathrm{N})/(4 \hbar\lambda)$ is expressed in terms of the 
thickness $L_y$ of the AFM strip, the paramagnetic susceptibility $\chi_\mathrm{N}$ and spin-diffusion length $\lambda$ of the NM. 
The quantities $\vect{\Omega}_{n;q}$ and $f_{n;q}$ are given by the expressions
\begin{eqnarray}
\vect{\Omega}_{n;q}&=&\sum_{\alpha,\beta}\big(  \vect{F}_{\alpha\beta;nq}^{+ -}  + \vect{F}_{\alpha\beta;nq}^{- +}  + 2 \vect{F}_{\alpha\beta;nq}^{z z}   \big) , \\
f_{n;q}&=&\sqrt{\frac{\sqrt{(1+(\lambda q)^2)^2+(\omega_n(q)\tau )^2}-1-(\lambda q)^2}{2\left((1+(\lambda q)^2)^2+(\omega_n(q)\tau )^2\right)}} . \nonumber 
\end{eqnarray}
Here, $\alpha,\beta\in \{ N_{1D},N_{1D-1}, N_{1D-2} \}$ are the AFM's sublattice spins lying at the AFM/NM interface (see Fig.~\ref{Fig3}b), $\tau$ is the spin-flip relaxation time in the NM, $\vect{F}_{\alpha\beta;nq}^{\eta \tilde{\eta} } = \vect{\Lambda}_{\alpha n;q}^{\eta}\Gamma_{\beta n;q}^{\tilde{\eta},*}$, where $\eta, \tilde{\eta}\in\{\pm,z\}$,  and the $\Lambda$ and $\Gamma$ tensors are defined in terms of the paraunitary matrix 
$\mathbf{T}_q^{-1}=[ \mathbf{U}_{q} , \mathbf{V}_{-q}^* ;   \mathbf{V}_{q} , \mathbf{U}_{-q}^*]$ that diagonalizes $\mathcal{H}_\text{AF}^{\rm (1D)}$:
\begin{eqnarray}
\vect{\Lambda}_{\alpha n;q}^\eta &=& i\tilde{J}_\mathrm{sd}(\vect{r}_\alpha^{-\eta}U_{\alpha n,q}+\vect{r}_\alpha^{+\eta}V_{\alpha n,q})\text{, }  \nonumber \\
\Gamma_{\beta n;q}^{\eta,*} &=& \frac{\tilde{J}_\mathrm{sd}}{2}(c_\beta^{-\eta}V_{\beta n,q}^*+c_\beta^{+\eta}U_{\beta n,q}^*) .\nonumber
\end{eqnarray}
In the above expressions, $\tilde{J}_\mathrm{sd}=\hbar N_sJ_\mathrm{sd} \sqrt{S/2N_\mathrm{N}N_\mathrm{AF}} $, where $N_{N}$ ($N_{AF}$) is the number of 
unit cells in the NM (1D AFM) and $N_s$ is the number of spins of one type (A, B or C) at the AFM/NM interface.
Furthermore, we have that $\vect{r}_\alpha^{\pm\pm}=\vect{r}_\alpha^\pm\times\vect{r}^\pm$, $\vect{r}_\alpha^{\pm z}=\vect{r}_\alpha^\pm\times\hat{\vect{z}}$, 
$c_\alpha^{\pm\pm}=\vect{r}_\alpha^\pm\cdot\vect{r}^\pm$, and  $c_\alpha^{\pm z}=\vect{r}_\alpha^\pm\cdot\hat{\vect{z}}$, where 
$\vect{r}_{\alpha}^\pm=\hat{\vect{x}}_{\alpha}\pm i\hat{\vect{y}}_{\alpha}$  and $\vect{r}^\pm=\hat{\vect{x}}\pm i\hat{\vect{y}}$.

Fig.~\ref{Fig3}c illustrates the relative contribution of first-gap edge states to the total spin current as a function of the AFM temperature for three in-plane DMI values: 3.5, 5, and 7 meV/$\hbar^2$. We focus on the $z$-component, which constitutes the dominant portion of the spin current~\cite{Comment1}. The temperature difference across the AFM/NM interface is fixed at  $T_\mathrm{AF}-T_\mathrm{N}=1$~K. The material parameters for the NM layer are set to $\lambda=70$~Å and $\tau=1$~psec~\cite{Adachi:prb2011}. In all three cases, the edge state contributions are substantial, with peak values ranging from 27\% to 31\%. These maxima occur at relatively low temperatures (30–50 K) due to the inflection point of $\coth\frac{\varepsilon_n(q)}{2k_BT_\mathrm{AF}}-\coth\frac{\varepsilon_n(q)}{2k_BT_\mathrm{N}}$ at $k_B T_{\rm AF}\sim\varepsilon_n/4.5$. Beyond these peaks, the contributions remain considerable, reaching 21–28\% at room temperature. The significant contribution to the spin current is characteristic of topological magnons and arises from their strong coupling to the itinerant charge carriers at the AFM/NM interface. This is demonstrated in Fig.~\ref{Fig3}d, which shows the contributions from all bands for $D_\parallel=5$ meV/$\hbar^2$ at 300 K. It is noteworthy that the mid-range contribution from the trivial magnon states is about  5\%. 
Thus, the contribution from the topological magnon band is 4–5 times higher than that of the trivial bands, underscoring the potential of edge states to greatly amplify the antiferromagnetic SSE. 
It should be emphasized that while edge states are typically associated with topological magnon bands, non-topological edge magnons are also expected to produce a similar enhancement of the SSE.

Theoretical investigations have demonstrated that the population of edge modes can be tuned and enhanced using a driving electromagnetic field~\cite{Malz:NatCom2019, Bhowmick:prb2023}. 
This effect provides a promising mechanism for exploring the influence of topological magnons on the SSE and could potentially be utilized to amplify thermal spin pumping.

In our investigations, we have disregarded the effects of inelastic relaxation mechanisms -- such as magnon-magnon and magnon-phonon interactions -- on the robustness of the topological magnon states. These relaxation mechanisms could influence the SSE, particularly at high temperatures. It is generally believed that the topological states remain intact as long as the linewidth of the magnon excitations is smaller than the gap between neighboring magnon bands. However, a recent theoretical study demonstrates that the edge states remain robust even when the damping rates are large enough to close the gap, thus indicating that the topological magnons exhibit resilience against substantial inelastic relaxation mechanisms~\cite{Hu:arXiv24}.     

Our study focused on a 2D sample where the charge carriers couple to the AFM spins along a line. However, we anticipate that the impact of topological magnons could be even more pronounced in 3D systems, where the magnons and charge carriers interact across a 2D AFM/NM interface. Recent experiments have already reported evidence of topological magnons in 3D AFMs~\cite{Yao:NatPhys2018,Bao:NatComm2018}, making it an exciting prospect for future research.

\section{Acknowledgements}
We thank Mathias Kläui for stimulating discussions.
KMDH acknowledges funding from the Research Council of Norway via Project No. 334202.


\appendix
\section{The tilting angle $\theta$}\label{ap:theta}
To minimize the energy of the spin Hamiltonian $\mathcal{H}_\text{AF}$ in Eq.~\eqref{eq:ham} in the main text with respect to the tilting angle $\theta$, we substitute the following ansatz for the spins into Eq.~\eqref{eq:ham}: 
$\vect{S}_i= \hbar S[ \cos \theta \hat{\vect{n}}_i + \sin \theta \hat{\vect{z}}]$. $\mathcal{H}_\text{AF}$ then becomes  
\begin{equation}
\begin{split}
\mathcal{H}_\text{AF} =&3\big(J(3\sin^2\theta-1)+(K_\perp\sin^2\theta-K\cos^2\theta)\\
&+\sqrt{3}(D_\perp\cos^2\theta+D_\parallel\sin2\theta)\big)N_{\rm uc} S^2\hbar^2 \text{,}
\end{split}
 \label{eq:unit}
\end{equation}
where $N_{\rm uc}$ is the number of primitive unit cells of the kagome AFM.
Differentiation with respect to $\theta$, yields the expression for the tilting angle in the ground state:
\begin{equation}
\theta=-\frac{1}{2}\arctan\left( \frac{2\sqrt{3}D_\parallel}{3J+K_\perp+K-\sqrt{3}D_\perp}\right) \text{.}
 \label{eq:theta}
\end{equation}
As expected, $\theta$ vanishes in the limit $D_\parallel\rightarrow 0$ of no in-plane DMI.

\section{The BdG Hamiltonian}\label{app:bdg}
The Bogoliubov-de-Gennes (BdG) matrix $\mathbf{H}_{\vect{k}}$ in Eq.~\eqref{eq:bdg} in the main text is given by
\begin{equation}\label{eq:matrix}
    \mathbf{H}_{\vect{k}}=
    \begin{pmatrix}
    \Gamma & \Lambda_\vect{a}^*& \Lambda_\vect{c} & \tilde\Gamma & \tilde\Lambda_\vect{a} & \tilde\Lambda_\vect{c}\\
    \Lambda_\vect{a} & \Gamma & \Lambda_\vect{b}^* & \tilde\Lambda_\vect{a} & \tilde\Gamma & \tilde\Lambda_\vect{b}\\
    \Lambda_\vect{c}^* & \Lambda_\vect{b} & \Gamma & \tilde\Lambda_\vect{c} & \tilde\Lambda_\vect{b} & \tilde\Gamma\\
    \tilde\Gamma & \tilde\Lambda_{-\vect{a}} & \tilde\Lambda_{-\vect{c}} & \Gamma & \Lambda_{-\vect{a}}& \Lambda_{-\vect{c}}^*\\
    \tilde\Lambda_{-\vect{a}} & \tilde\Gamma & \tilde\Lambda_{-\vect{b}} & \Lambda_{-\vect{a}}^* & \Gamma & \Lambda_{-\vect{b}}\\
    \tilde\Lambda_{-\vect{c}} & \tilde\Lambda_{-\vect{b}} & \tilde\Gamma & \Lambda_{-\vect{c}} & \Lambda_{-\vect{b}}^* & \Gamma
    \end{pmatrix}
    \text{,}
    \end{equation}
where
\begin{eqnarray}\label{eq:whole}
&&\overset{(\sim)}{\Lambda}_\vect{m}=\overset{(\sim)}{\Lambda}\cos(\vect{k}\cdot\vect{m})=\overset{(\sim)}{\Lambda}_{-\vect{m}},\quad\vect{m}\in\{\vect{a},\vect{b},\vect{c}\}\nonumber\\
&&\Gamma= \hbar^2 S\big((2J+K_\perp)(1-3\sin^2\theta)+K(2-3\sin^2\theta)\nonumber\\
&&\qquad\qquad-2\sqrt{3}(D_\perp\cos^2\theta+D_\parallel\sin2\theta)\big), \nonumber\\
&&\tilde\Gamma=-S\hbar^2(K_\perp\cos^2\theta-K\sin^2\theta)\text{,}\nonumber\\
&&\Lambda =S\hbar^2\Bigg(-J\left(\sin^2\theta-\frac{\cos^2\theta}{2}+i\sqrt{3}\sin\theta\right)\\
&&\qquad\qquad+D_\perp\left(\frac{\sqrt{3}}{2}+\frac{\sqrt{3}}{2}\sin^2\theta-i\sin\theta\right)\nonumber\\
&&\qquad\qquad-D_\parallel\left(\frac{\sqrt{3}}{2}\sin2\theta+i\cos\theta\right)\Bigg)\text{,}\nonumber\\
&&\tilde\Lambda =-\hbar^{2}\frac{\sqrt{3}}{2}S\left((\sqrt{3}J-D_\perp)\cos^2\theta-D_\parallel\sin2\theta\right)\text\nonumber{,}
\end{eqnarray}
and the asterisk (*) denotes complex conjugation.

\section{Derivation of the spin polarization}\label{app:sp}
In this section, we derive the formula for the average spin polarization carried by a given spin-wave mode.
The BdG Hamiltonian in Eq.~\eqref{eq:bdg} of the main text is diagonalized by a paraunitary matrix~\cite{Colpa}:
\begin{equation}\label{eq:uv}
    T_{\vect{k}}^{-1}=
	\begin{pmatrix}
       U_{\vect{k}} & V_{-\vect{k}}^*\\
       V_{\vect{k}} & U_{-\vect{k}}^*
	\end{pmatrix}\text{.}
    \end{equation}
Consequently, the relationship between the operators $(\vect{\alpha}_{\vect{k}}^\dagger,\vect{\alpha}_{-\vect{k}})$ describing the sublattice excitations and the ones characterizing the elementary magnon excitations,  $(\vect{\gamma}_{\vect{k}}^\dagger,\vect{\gamma}_{-\vect{k}})$, becomes     
\begin{equation}\label{eq:new}
\begin{split}
    &\alpha_{\lambda,\vect{k}}= \sum_j U_{\lambda j,\vect{k}}\gamma_{j,\vect{k}}+V_{\lambda j,-\vect{k}}^*\gamma_{j,-\vect{k}}^\dagger\quad\text{and}\\
    &\alpha_{\lambda,\vect{k}}^\dagger= \sum_j U_{\lambda j,\vect{k}}^*\gamma_{j,\vect{k}}^\dagger+V_{\lambda j,-\vect{k}}\gamma_{j,-\vect{k}}\text{,}
\end{split}
    \end{equation}
where the Greek index $\lambda$ enumerates the sublattice sites and the Latin index $j$ enumerates the spin-wave energy bands. 
Thus, the number operator (of each sublattice) expressed in terms of the operators $(\vect{\gamma}_{\vect{k}}^\dagger,\vect{\gamma}_{-\vect{k}})$ is given by 
\begin{eqnarray}\label{eq:number}
&&\sum_{\vect{k}\in\mathrm{1BZ}}\alpha_{\lambda,\vect{k}}^\dagger\alpha_{\lambda,\vect{k}} =  \sum_{\vect{k},i,j}\big(   \delta_{ij}( V_{\lambda j,\vect{k}}V_{i\lambda,\vect{k}}^\dagger)   +  \\
&&U_{\lambda i,\vect{k}}^*V_{\lambda j,-\vect{k}}^*\gamma_{i,\vect{k}}^\dagger\gamma_{j,-\vect{k}}^\dagger+V_{\lambda i,-\vect{k}}U_{\lambda j,\vect{k}}\gamma_{i-\vect{k}}\gamma_{j,\vect{k}} + \nonumber \\
&& (U_{i\lambda,\vect{k}}^\dagger U_{\lambda j,\vect{k}}+V_{i\lambda,\vect{k}}^\dagger V_{\lambda j,\vect{k}})\gamma_{i,\vect{k}}^\dagger\gamma_{,j,\vect{k}} \big) \text{,}\nonumber
\end{eqnarray}
where we have utilized that $\gamma_{\vect{k},j}\gamma_{\vect{k},i}^\dagger=\delta_{ij}+\gamma_{i,\vect{k}}^\dagger\gamma_{j,\vect{k}}$. 
The momenta are summed over the first Brillouin zone (1BZ). From Eq.~\eqref{eq:number}, we find an expression for the average of the spin polarization in the vacuum state $\ket{0}$:
\begin{eqnarray}\label{eq:tot}
\bra{0}\vect{S}_\mathrm{tot}\ket{0}&=&\bra{0}\sum_{\lambda}\hbar\left( N_{\rm uc} S-\sum_{\vect{k}}\alpha_{\lambda,\vect{k}}^\dagger\alpha_{\lambda,\vect{k}}\right)\hat{\vect{z}}_\lambda\ket{0} \nonumber \\
&=&\sum_{\lambda}\hbar\left(N_\mathrm{uc}S-\sum_{\vect{k}}(V_{\vect{k}}V_{\vect{k}}^\dagger )_{\lambda\lambda}\right)\hat{\vect{z}}_\lambda\text{,}
\end{eqnarray}
where $\hat{\vect{z}}_\lambda$ is the local $z$-direction of the spin at sublattice site $\lambda$ and the summation is performed over the three lattice sites in the unit cell $\lambda\in\{A,B,C\}$. The second term in Eq.~\eqref{eq:tot}, proportional to $(V_{\vect{k}}V_{\vect{k}}^\dagger )_{\lambda\lambda}$, originates from the quantum fluctuations in the ground state.
Similarly, we can evaluate the average of the total spin polarization of a spin system in which a single magnon state $\ket{n,\vect{q}}=\gamma_{n,\vect{q}}^\dagger\ket{0}$ is excited
\begin{eqnarray}\label{eq:totus}
&&\bra{n,\vect{q}}\vect{S}_\mathrm{tot}\ket{n,\vect{q}}\\
&&=\bra{0}\gamma_{n,\vect{q}}\sum_{\lambda}\hbar\left(N_\mathrm{uc}S-\sum_{\vect{k}}\alpha_{\lambda,\vect{k}}^\dagger\alpha_{\lambda,\vect{k}}\right)\hat{\vect{z}}_\lambda\gamma_{n,\vect{q}}^\dagger\ket{0}\nonumber\\
&&= \bra{0}\vect{S}_\mathrm{tot}\ket{0}-\hbar\sum_{\lambda}(\abs{U_{\lambda n,\vect{q}}}^2+\abs{V_{\lambda n,\vect{q}}}^2)\hat{\vect{z}}_\lambda\text{.}\nonumber
\end{eqnarray}
The average spin carried by a single magnon is thus given by
\begin{equation}\label{eq:spinpol}
\begin{split}
\langle\vect{S}_{\mathrm{SW};n,\vect{k}}\rangle &= -(\bra{n,\vect{k}}\vect{S}_\mathrm{tot}\ket{n,\vect{k}}-\bra{0}\vect{S}_\mathrm{tot}\ket{0})\\
&=\hbar\sum_{\lambda}(\abs{U_{\lambda n,\vect{k}}}^2+\abs{V_{\lambda n,\vect{k}}}^2)\hat{\vect{z}}_\lambda \text{.}
\end{split}
\end{equation}

\section{Expression for the SSE}\label{app:model}
Below, we derive a microscopic expression for the SSE of a kagome AFM/NM heterostructure. Similar derivations have been carried out for heterostructures of ferromagnets and collinear AFMs~\cite{Adachi:prb2011, Ohnuma:prb2013}.

\subsection{The model for the AFM/NM heterostructure}
The Hamiltonian describing the heterostructure consisting of a quasi-1D AFM coupled to a 2D NM is
\begin{equation}\label{eq:totall}
\mathcal{H}=\mathcal{H}_\mathrm{NM} + \mathcal{H}_\mathrm{AF}^{\rm (1D)} + \mathcal{H}_\mathrm{I}\text{.}
\end{equation}
Here, the Hamiltonian of the NM is given by
\begin{equation}\label{eq:normal}
\mathcal{H}_\mathrm{NM}=\sum_{\vect{k},\vect{q}}\mathbf{c}_{\vect{k}}^\dagger \left(\varepsilon_{\vect{k}}\delta_{\vect{k},\vect{q}}+U_{\vect{k}-\vect{q}}\left(1+i \alpha_\mathrm{so}\vect{\sigma}\cdot(\vect{k}\times\vect{q})\right)\right)\mathbf{c}_{\vect{q}}\text{,}
\end{equation}
where $\mathbf{c}_{\vect{k}}^{\dagger} = [c_{\vect{k},\uparrow}^{\dagger} , c_{\vect{k},\downarrow}^{\dagger}]$ is the spinor containing the creation operators 
$c_{\vect{k},\sigma}^{\dagger}= (1/\sqrt{N_N})\sum_{\vect{r}_i\in N}c_{i,\sigma}^{\dagger}e^{i\vect{k}\cdot\vect{r}_i}$
of electron states with momentum $\vect{k}$  and spin $\sigma\in \{ \uparrow,\downarrow \}$. $U_{\vect{k}-\vect{q}}$ is the Fourier transform of the scalar impurity potential, $\alpha_{\rm so}$ parametrizes the SOC strength at the impurities, and $\varepsilon_{\vect{k}}$ represents the energy of the free electrons.
The Hamiltonian of the AFM is
\begin{equation}\label{eq:bdgigjen}
    \mathcal{H}_\mathrm{AF}^{\rm (1D)}=\frac{1}{2}\sum_{q_x}(\vect{\alpha}_{q_x}^\dagger,\vect{\alpha}_{-q_x})\mathbf{H}_{q_x}
    \begin{pmatrix}
    \vect{\alpha}_{q_x}\\
    \vect{\alpha}_{-q_x}^\dagger
    \end{pmatrix}
    \text{.}
    \end{equation}
Here, $\vect{\alpha}_{q_x}=(a_{1,q_x},a_{2,q_x},\dots,a_{N_{\rm 1D},q_x})^T$ where $N_{\rm 1D}=3N_\mathrm{y}-1$ is the number of spin sublattices within the 1D unit cell with $N_\mathrm{y}$ being the number of primitive unit cells along the $y$-axis. 
The Fourier transformed of the magnon operators is defined by
$a_{\alpha, q_x}^{\dagger}= (1/\sqrt{N_{AF}})\sum_{\vect{r}_i\in AF}a_{\alpha, i}^{\dagger}e^{i q_x \hat{\vect{x}}\cdot\vect{r}_i}$.
The operators characterizing the spin waves are $\vect{\gamma}_{q_x}=(\gamma_{1,q_x},\gamma_{2,q_x},\dots,\gamma_{N_{\rm 1D},q_x})^T$ and are related to the $\vect{\alpha}_{q_x}$-basis via the $2N_{\rm1D}\times 2N_{\rm1D}$ paraunitary matrix 
$\mathbf{T}_{q_x}^{-1} $ 
diagonalizing $\mathbf{H}_{q_x}$ in Eq.~\eqref{eq:bdgigjen}:
\begin{equation}\label{eq:novo}
    a_{\alpha,q_x}=\sum_j (U_{\alpha j,q_x}\gamma_{j,q_x}+V_{\alpha j,-q_x}^*\gamma_{j,-q_x}^\dagger ) .
\end{equation}
Here, $\mathbf{U}_{q_x}$ and $\mathbf{V}_{q_x}$ are the $N_{\rm 1D}\times N_{\rm 1D}$-dimensional submatrices of $\mathbf{T}_{q_x}^{-1}=  [\mathbf{U}_{q_x},\mathbf{V}_{-q_x}^\ast; \mathbf{V}_{q_x},\mathbf{U}_{-q_x}^\ast] $.

To linear order in the magnon operators, the Fourier transform of the Hamiltonian $\mathcal{H}_\mathrm{I}$ characterizing the interfacial exchange coupling between the antiferromagnetic magnons and the charge carriers of the NM is 
\begin{equation}\label{eq:sdfull}
\begin{split}
\mathcal{H}_\mathrm{I}=\frac{1}{\sqrt{N_\mathrm{N}N_\mathrm{AF}}}\sum_{\vect{k},q_x ,\alpha\in E}  &J_{\mathrm{sd}, \vect{k}- q_x \hat{\vect{x}}}\big((\vect{s}_{\vect{k}}\cdot\hat{\vect{x}}_\alpha)S_{\alpha,q_x}^x\\
&\qquad+(\vect{s}_{\vect{k}}\cdot\hat{\vect{y}}_\alpha)S_{\alpha,q_x}^y\big)  \text{.}
\end{split}
\end{equation}
Here, the Fourier transformed exchange coupling is $J_{\mathrm{sd},\vect{q}}=\sum_{\vect{r}_i\in I}J_{\mathrm{sd},i} e^{-i\vect{q}\cdot\vect{r}_i} = N_s J_\mathrm{sd} \delta_{q_x 0}$,  
$\delta_{ij}$ is the Kronecker-delta, $N_N$ ($N_{AF}$)  is the number of unit cells of the NM (quasi-1D AFM), $N_s$ is the number of spins of one type ($A$, $B$
and $C$) at the interface, and $E\in \{ N_{1D}, N_{1D} -1, N_{1D} -2  \}$ is the set of sublattice spins within the unit cell of the quasi-1D AFM that couple to the itinerant charge carriers of the NM.
The term proportional to $S_{\alpha,\vect{q}}^z$ is disregarded as it is second order in the magnon operators. 
The Fourier transformed spin density, $\vect{s}_{\vect{q}}= (1/\sqrt{N_N}) \sum_{\vect{r}_i\in N} \vect{s}_i \exp (i \vect{q} \cdot \vect{r}_i )$,  in the NM is 
\begin{equation}\label{eq:spindensityk}
\vect{s}_{\vect{q}}=\frac{\hbar}{2}\frac{1}{\sqrt{N_N}}\sum_{\vect{k}, \tau, \tau^{'}}c_{\vect{k}+\vect{q},\tau}^\dagger\vect{\sigma}_{\tau\tau^{'}}c_{\vect{k},\tau^{'}}\text{,}
\end{equation}
whereas the $S_{\alpha,q_x}^x$ and $S_{\alpha,q_x}^y$ operators are given by
 \begin{equation}\label{eq:newdef}
\begin{split}
&S_{\alpha,q_x}^x=\hbar\sqrt{\frac{S}{2}}(a_{\alpha,q_x}+a_{\alpha,-q_x}^\dagger) \quad\text{and}\\
& S_{\alpha,q_x}^y=\frac{\hbar}{i}\sqrt{\frac{S}{2}}(a_{\alpha,q_x}-a_{\alpha,-q_x}^\dagger) \text{.}
\end{split}
\end{equation}
Here, $ S_{\alpha,q_x}^j = (1/\sqrt{N_{AF}})\sum_{\vect{r}_i\in AF} S_{\alpha, i}^j e^{-i q_x \hat{\vect{x}}\cdot \vect{r}_i }$ where $\vect{r}_i$ runs over the lattice sites and $\alpha\in \{1, ..., N_{1D} \}$ labels the sublattice spins within the unit cell of the quasi-1D AFM.   
By using Eqs.~\eqref{eq:novo} and \eqref{eq:newdef}, along with $J_{\mathrm{sd}, \vect{k} - q_x\hat{\vect{x}}} =N_s J_\mathrm{sd}  \delta_{ k_x q_x}$,
the interfacial Hamiltonian \eqref{eq:sdfull} can be expressed in terms of the magnon operators as
\begin{equation}\label{eq:sdexgammasum}
\begin{split}
\mathcal{H}_\mathrm{I} = 
\sum_{\vect{k},q_x,j, \alpha\in E}\big( & \Gamma_{\alpha j;q_x}^-\gamma_{j,q_x}s_{\vect{k}}^-   +\Gamma_{\alpha j;q_x}^+\gamma_{j,q_x}s_{\vect{k}}^+ \\
&+ 2 \Gamma_{\alpha j;q_x}^z\gamma_{j,q_x}s_{\vect{k}}^z+ h.c. \big)\delta_{k_x q_x}
\text{.}
\end{split}
\end{equation}
Here, $s_{\vect{q}}^{\pm}= (s_{\vect{q}}^x\pm is_{\vect{q}}^y)$, $j\in \{1, ..., N_{\rm 1D} \}$, and the $\Gamma$-coefficients are defined in the main text.

\subsection{The spin current }
The total spin accumulation $\vect{s}_\mathrm{Tot}=(\hbar/2)\sum_{i}\mathbf{c}_{i}^\dagger\boldsymbol{\sigma} \mathbf{c}_{i}$ in the NM region can be expressed as
\begin{equation}\label{eq:totspin}
s_\mathrm{Tot}^i=\frac{\hbar}{2}\sum_{\vect{k}, \tau, \tau^{'}}c_{\vect{k},\tau}^\dagger \sigma_{\tau\tau^{'}}^ic_{\vect{k},\tau^{'}}=\lim_{\vect{p}\rightarrow0}\sqrt{N_\mathrm{N}}s_\vect{p}^i .
\end{equation}
By using the above expression along with the Heisenberg equation, the spin current pumped into the NM region due to the interfacial exchange coupling to the AFM can be written as
\begin{eqnarray}\label{Eq:Is1}
\vect{I}_\mathrm{s} &=& \lim_{\vect{p}\rightarrow0} \sqrt{N_\mathrm{N}} \frac{i}{\hbar} \langle [\mathcal{H}_\mathrm{I},\vect{s}_\vect{p}] \rangle\\
&=&\frac{1}{\sqrt{N_\mathrm{N}N_\mathrm{AF}}}\sum_{\vect{k},q_x,\alpha\in E}J_{\mathrm{sd},\vect{k}- q_x\hat{\vect{x}}}\nonumber \\
&&\times\langle \left((\hat{\vect{x}}_\alpha\times\vect{s}_\vect{k}) S_{\alpha,q_x}^x+(\hat{\vect{y}}_\alpha\times\vect{s}_\vect{k}) S_{\alpha,q_x}^y\right) \rangle \text{,} \nonumber
\end{eqnarray}
where we have used that  $(i/\hbar)[s_\vect{k}^i,s_\vect{q}^j]= (\epsilon_{jik}/\sqrt{N_\mathrm{N}}) s_{\vect{k}+\vect{q}}^k$.
Substitution of Eqs.~\eqref{eq:spindensityk}-\eqref{eq:newdef} into Eq.~\eqref{Eq:Is1} yields
\begin{eqnarray}\label{eq:scnewesum}
\vect{I}_\mathrm{s}&=&\sum_{\vect{k},q_x,j, \alpha\in E}\mathfrak{Re}\big(\vect{\Lambda}_{\alpha j;q_x}^+C_{j,q_x,\vect{k}}^{-,<}(t,t)+\vect{\Lambda}_{\alpha j;q_x}^-C_{j,q_x,\vect{k}}^{+,<}(t,t) \nonumber \\
&&\qquad\qquad\qquad+2\vect{\Lambda}_{\alpha j;q_x}^zC_{j,q_x,\vect{k}}^{z,<}(t,t)\big) \delta_{k_x q_x}\text{,}
\end{eqnarray}
after expressing the $\vect{\alpha}_{k_x}$-operators in terms of the magnon operators using Eq.~\eqref{eq:novo}. Here,
\begin{equation}\label{eq:cors}
\begin{split}
&C_{j,q_x,\vect{k}}^{-,<}(t',t)=-i\langle s_\vect{k}^-(t)\gamma_{j,q_x}(t')\rangle\text{,}\\
&C_{j,q_x,\vect{k}}^{+,<}(t',t)=-i\langle s_\vect{k}^+(t)\gamma_{j,q_x}(t')\rangle\text{,}\\
&C_{j,q_x,\vect{k}}^{z,<}(t',t)=-i\langle s_\vect{k}^z(t)\gamma_{j,q_x}(t')\rangle
\end{split}
\end{equation}
are the three lesser Green's functions expressing the correlations between the magnons and the charge carriers. 
The $\vect{\Lambda}$-vectors are defined in the main text. The operators are defined in the Heisenberg picture with respect to $\mathcal{H}$ in Eq.~\eqref{eq:totall} and $\langle ... \rangle = (1/Z) {\rm Tr} [\exp (-\mathcal{H}/k_BT) ...]$.
Furthermore, the lesser Green's functions can be expressed by the Keldysh ($C^{\eta, K}$), retarded ($C^{\eta, R}$), and advanced ($C^{\eta,A}$) Green's functions via the equation ($\eta\in \{ +,-,z  \}$)~\cite{Rammer:book}
\begin{equation}\label{eq:lessKRA}
C_{j,q_x,\vect{k}}^{\eta, <}(\omega)=\frac{1}{2}\left(C_{j,q_x,\vect{k}}^{\eta, K}(\omega)-C_{j,q_x,\vect{k}}^{\eta, R}(\omega)+C_{j,q_x,\vect{k}}^{\eta, A}(\omega)\right)\text{,}
\end{equation}
where the Fourier transformation of the Green's functions is defined by
\begin{equation}\label{eq:fourierless}
C_{j,q_x,\vect{k}}(t',t)=\frac{1}{2\pi}\int_{-\infty}^{\infty} d\omega C_{j,q_x,\vect{k}}(\omega)e^{-i\omega(t'-t)}\text{.}
\end{equation}
In Eq.~\eqref{eq:lessKRA}, the temperature dependence of the Green's function emerges through the Keldysh component via the fluctuation-dissipation theorem, in alignment with results for ferromagnets, ferrimagnets, and collinear antiferromagnets~\cite{Adachi:prb2011,Ohnuma:prb2013}. 
In contrast, the retarded and advanced Green's functions contain information about the system's spectral properties.
Thus, inserting the expressions in Eq.~\eqref{eq:fourierless} and Eq.~\eqref{eq:lessKRA} into Eq.~\eqref{eq:scnewesum}, we arrive at:
\begin{equation}\label{eq:scnewesumK}
\begin{split}
&\vect{I}_\mathrm{s}=\sum_{\vect{k},q_x, j, \alpha\in E}\int_{-\infty}^{\infty}\frac{d\omega}{4\pi}\mathfrak{Re}\big(\vect{\Lambda}_{\alpha j;q_x}^+ C_{j,q_x,\vect{k}}^{-,K}(\omega)\\
&+\vect{\Lambda}_{\alpha j;q_x}^- C_{j,q_x,\vect{k}}^{+,K}(\omega)+2\vect{\Lambda}_{\alpha j;q_x}^zC_{j,q_x,\vect{k}}^{z,K}(\omega)\big) \delta_{k_x q_x}
\text{.}
\end{split}
\end{equation}
Below, we use linear response theory to derive expressions for the Keldysh Green's functions to first order in $\mathcal{H}_\mathrm{I}$.

\subsection{Linear response calculation}
To calculate the Keldysh Green's functions in Eq.~\eqref{eq:scnewesumK}, we consider the contour-ordered Green’s function along the Keldysh contour $\mathcal{C}$~\cite{Rammer:book}:
\begin{equation}\label{eq:cor1}
C_{j,q_x,\vect{k}}^{\eta,(\tau,\tau')}(t',t) = -i\langle \mathcal{T}_\mathcal{C}e^{-\frac{i}{\hbar}\int_\mathcal{C}\mathcal{H}_\mathrm{I}(t_1)dt_1}\gamma_{j,q_x}(t',\tau')s_\vect{k}^{\eta}(t,\tau)\rangle_0 ,
\end{equation}
where $\eta\in \{ +, -, z  \}$, the operators are defined in the interaction picture, and the average $\langle ... \rangle_0$ is evaluated in the non-interacting regime where $\mathcal{H}_\mathrm{I}=0$.  
$\mathcal{T}_\mathcal{C}$ is the time-ordering operator along the Keldysh contour with upper/lower branch labeled by the indices $\tau,\tau'$. 
Using the expression~\eqref{eq:sdexgammasum} for $\mathcal{H}_\mathrm{I}$ in Eq.~\eqref{eq:cor1}, followed by an expansion of the exponent to first order in $\mathcal{H}_\mathrm{I}$ and the application of Wick's theorem, one finds
\begin{equation}\label{eq:corzomega1}
\underline{C^\eta}_{j,q_x,\vect{k}}(\omega)=\sum_{\alpha\in E}\frac{1}{\hbar}\Gamma_{\alpha j;q_x}^{\eta ,*}\underline{G_\mathrm{AF}}_{j,q_x}(\omega)\underline{G_\mathrm{N}}_{\vect{k}}(\omega) \delta_{k_x q_x}  \text{.}
\end{equation} 
Here, the underlined quantities represent 2 × 2 Green’s function in Keldysh space, $\underline{G}=[ G^R,  G^K ;  0 , G^A ]$, $\underline{G_\mathrm{AF}}_{j,q_x}(\omega)$ is the Green's function describing the isolated magnon system, and  
$\underline{G_\mathrm{N}}_{\vect{k}}(\omega)$ characterizes the spin-density correlations in the NM.
Substituting Eq.~\eqref{eq:corzomega1} into  Eq.~\eqref{eq:scnewesumK}, we obtain the expression
\begin{equation}\label{eq:grand}
\vect{I}_\mathrm{s}=\sum_{\vect{k},q_x ,j}\int_{-\infty}^\infty\frac{d\omega}{4\pi\hbar}\vect{\Omega}_{j; q_x} \delta_{k_x q_x} \mathfrak{Re}\left(\underline{G_\mathrm{AF}}_{j,q_x}(\omega)\underline{G_\mathrm{N}}_{\vect{k}}(\omega)\right)_{12}  
\end{equation}
where
\begin{equation}\label{eq:grandomega}
\vect{\Omega}_{j;q_x}=\sum_{\alpha,\beta\in E }\vect{\Lambda}_{\alpha j;q_x}^+\Gamma_{\beta j;q_x}^{-,*}+\vect{\Lambda}_{\alpha j;q_x}^-\Gamma_{\beta j;q_x}^{+,*}
+2\vect{\Lambda}_{\alpha j;q_x}^z\Gamma_{\beta j;q_x}^{z,*} \text{.}
\end{equation}
Using the bosonic fluctuation-dissipation theorem $G^{K} (\omega)= [G^{R} (\omega) - G^{A} (\omega)] \coth (\beta \hbar \omega/2)$ (where $\beta= 1/k_B T$), the real part of the Keldysh component of the Green's function product can be expressed as 
\begin{equation}\label{eq:real}
\begin{split}
&\mathfrak{Re}\left(\underline{G_\mathrm{AF}}_{j,q_x}(\omega)\underline{G_\mathrm{N}}_{\vect{k}}(\omega)\right)_{12}\\
&=\mathfrak{Re}\big(G_{\mathrm{AF},j,q_x}^R(\omega)G_{\mathrm{N},\vect{k}}^K(\omega) + G_{\mathrm{AF},j,q_x}^K(\omega)G_{\mathrm{N},\vect{k}}^A(\omega)\big) \\
&=2\mathfrak{Im}G_{\mathrm{AF},j,q_x}^R(\omega)\mathfrak{Im}G_{\mathrm{N},\vect{k}}^R(\omega) \mathcal{F} (\omega, T_\mathrm{AF}, T_\mathrm{N} ) , 
\end{split}
\end{equation}
where
\begin{equation}
\mathcal{F} (\omega, T_\mathrm{AF}, T_\mathrm{N} ) = \coth\frac{\hbar\omega}{2k_BT_\mathrm{AF}}-\coth\frac{\hbar\omega}{2k_BT_\mathrm{N}} .
\end{equation}
Here, $T_\mathrm{AF}$ ($T_\mathrm{N}$) is the temperature of the AFM (NM). The retarded Green's function of the spin-density correlations (i.e., the spin susceptibility) in the NM is given by~\cite{Adachi:prb2011, Ohnuma:prb2013}
\begin{equation}\label{eq:normalmetal}
\begin{split}
&G_{\mathrm{N},\vect{k}}^R(\omega)=\frac{\chi_\mathrm{N}}{1+\lambda^2\vect{k}^2-i\omega\tau}\quad\text{and}\\
& \mathfrak{Im}G_{\mathrm{N},\vect{k}}^R(\omega)=\frac{\chi_\mathrm{N}\omega\tau}{(1+\lambda^2\vect{k}^2)^2+(\omega\tau)^2}\text{,}
\end{split}
\end{equation}
where $\chi_\mathrm{N}$ denotes the paramagnetic susceptipility of the NM, $\lambda$ is the spin-diffusion length and $\tau$ is the spin-flip relaxation time. 
The spin-flip relaxation time arises from the SOC in the NM region. 
The retarded Green's function for the magnons in the AFM is 
\begin{equation}\label{eq:Antiferromagnet}
\begin{split}
&G_{\mathrm{AF},j,q_x}^R(\omega)=\frac{1}{\omega-\omega_j(q_x)+i\alpha_G\omega}\quad\text{and}\\
&\mathfrak{Im}G_{\mathrm{AF},j,q_x}^R(\omega)=\frac{-\alpha_G\omega}{(\omega-\omega_j(q_x))^2+(\alpha_G\omega)^2}\text{,}
\end{split}
\end{equation}
where $\hbar\omega_j(q_x)=\varepsilon_j(q_x)$ is the eigenvalue of the $j$th spin-wave mode with momentum $q_x$, and $\alpha_G$ is the Gilbert damping of the spins.
Substituting Eq.~\eqref{eq:real} into Eq.~\eqref{eq:grand}, we arrive at
\begin{eqnarray}\label{eq:grander}
\vect{I}_\mathrm{s}&=&\sum_{k_x=q_x,k_y,j}\int_{-\infty}^\infty\frac{d\omega}{2\pi\hbar}\vect{\Omega}_{j;q_x}\mathfrak{Im}G_{\mathrm{AF},j,q_x}^R(\omega)\mathfrak{Im}G_{\mathrm{N},\vect{k}}^R(\omega)\nonumber\\
&&\qquad\qquad\times\left(\coth\frac{\hbar\omega}{2k_BT_\mathrm{AF}}-\coth\frac{\hbar\omega}{2k_BT_\mathrm{N}}\right) \text{.}
\end{eqnarray}

Note that $\mathfrak{Im}G_{\mathrm{AF},j,q_x}^R(\omega)$ is strongly peaked at the frequency $\omega > 0$ corresponding to the eigenfrequency $\omega_j(q_x)$ of the magnon. 
We can, therefore, simplify the above result further by considering the low damping limit $\alpha_G \rightarrow 0$. In this limit, the imaginary part of the magnon Green's function can be approximated by
$\lim_{\alpha_G\rightarrow0}\mathfrak{Im}G_{\mathrm{AF},j,q_x}^R(\omega)\approx -\pi\delta(\omega-\omega_j(q_x))$.
Thus, we can perform the integration over $\omega$ in Eq.~\eqref{eq:grander}, which yields
\begin{equation}\label{eq:plusgrand}
\begin{split}
\lim_{\alpha_G\rightarrow0}\vect{I}_\mathrm{s}&=-\sum_{k_x=q_x,k_y,j}\frac{1}{2\hbar}\vect{\Omega}_{j;q_x}\mathfrak{Im}G_{\mathrm{N},\vect{k}}^R(\omega_j(q_x))\\
&\times\left(\coth\frac{\hbar\omega_j(q_x)}{2k_BT_\mathrm{AF}}-\coth\frac{\hbar\omega_j(q_x)}{2k_BT_\mathrm{N}}\right)\text{.}
\end{split}
\end{equation}

The quantity $\vect{\Omega}_{j;q_x}$ does not depend on $k_y$, which occurs only in the NM Green's function. 
Considering the continuum limit $a\rightarrow 0$, this allows us to integrate out the $k_y$-dependency of the NM Green's function
\begin{eqnarray}\label{eq:integrateky}
&\sum_{k_y}&\mathfrak{Im}G_{\mathrm{N},\vect{k}}^R(\omega_j(q_x)) \\
&=&\frac{L_y}{2\pi}\int_{-\infty}^\infty dk_y\frac{\chi_\mathrm{N}\omega_j(q_x)\tau}{(1+\lambda^2q_x^2+\lambda^2 k_y^2)^2+(\omega_j(q_x)\tau)^2} \nonumber \\
&=&\frac{L_y\chi_\mathrm{N}}{2\lambda}\sqrt{\frac{\sqrt{(1+(\lambda q_x)^2)^2+(\omega_j(q_x)\tau)^2}-1-(\lambda q_x)^2}{2\left((1+(\lambda q_x)^2)^2+(\omega_j(q_x)\tau)^2\right)}} \nonumber
\end{eqnarray}
where $L_y$ is the length of the NM sample in $y$-direction. 
Substituting the result from Eq.~\eqref{eq:integrateky} into Eq.~\eqref{eq:plusgrand} and renaming $q_x \rightarrow q$ and $j\rightarrow n$, we arrive at the final expression for the thermally-induced spin current in the NM region from Eq.~\eqref{eq:ultimate} in the main text.



\begin{thebibliography}{99}
\bibitem{Comment2}An external magnetic field along the $z$-axis increases the overall spin polarization of the magnon system (and, consequently, $I_{s,z}$) and potentially also drives the system into different topological regimes. However, the fundamental physics of the strong coupling between charge carriers and topological magnons remains unaffected.
\end{thebibliography}
\end{document}